\documentclass{elsartL}
\usepackage{graphicx}
\usepackage{amssymb}
\usepackage{amsmath}
\usepackage{physics}
\usepackage{mathrsfs}
\usepackage{mathtools}
\usepackage{appendix}
\newcommand{\avg}[1]{\left< #1 \right>} 
\begin{document}

\begin{frontmatter}
\title{Determination of the masses and decay widths of the well-established $s$ and $p$ baryon resonances below $2$ GeV}
\author{Evangelos Matsinos}

\begin{abstract}This study appertains to the extraction of estimates for the masses and for the partial decay widths to pion-nucleon final states of the well-established (four-star) $s$ and $p$ baryon resonances below $2$ 
GeV. These estimates are exclusively obtained from the data contained in the recent compilation by the Particle Data Group. Only the $N$ ($S=0,I=1/2$) and the $\Delta$ ($S=0,I=3/2$) states are considered, where $S$ and $I$ 
denote strangeness and total isospin, respectively.\\
\noindent {\it PACS 2010:} 13.30.-a, 14.20.Gk
%
\end{abstract}
\end{frontmatter}

\section{\label{sec:Introduction}Introduction}

This work is motivated by the application of its results in a pion-nucleon ($\pi N$) interaction model (ETH model), based on $s$-, $u$-, and $t$-channel Feynman graphs \cite{goudsmit1994,matsinos2017}, see Fig.~\ref{fig:FeynmanGraphsETHZ}. 
Regarding the $t$-channel contributions to the strong-interaction (hadronic) part of the $s$- and $p$-wave scattering amplitudes of this model, relevant (as intermediate states) are the scalar-isoscalar $I^G\,(J^{PC}) = 0^+\,(0^{++})$ 
and the vector-isovector $I^G\,(J^{PC}) = 1^+\,(1^{--})$ mesons~\footnote{In this work, $I$, $J$, $G$, $P$, $C$, and $S$ will denote total isospin, total angular momentum, G-parity, parity, charge conjugation, and strangeness, 
respectively.}. The results of an analysis of the available data, as they appear in the recent compilation \cite{pdg2020} by the Particle Data Group (PDG), can be found in Ref.~\cite{matsinos2020}.

This work can be thought of as a continuation of the analysis performed in Ref.~\cite{matsinos2020}. The goal here is the extraction from the data, listed in Ref.~\cite{pdg2020}, of estimates for the physical quantities (i.e., 
for the masses and for the partial decay widths to $\pi N$ final states) which are required for fixing the $s$- and $u$-channel contributions to the hadronic part of the $s$- and $p$-wave scattering amplitudes of the ETH 
model. Although the $s$- and $u$-channel graphs with $N$ and $\Delta (1232)$ intermediate states are dominant (see Fig.~\ref{fig:FeynmanGraphsETHZ}, lower part), contained in the ETH model since its introduction in the early 
1990s \cite{goudsmit1994} have also been the contributions originating from the exchanges of all other well-established (four-star) $N$ ($S=0,I=1/2$) and $\Delta$ ($S=0,I=3/2$) baryon resonances with rest masses below $2$ GeV, 
collectively classified as `higher baryon resonances (HBRs)' in all the publications of the ETH model (see cited works in Ref.~\cite{matsinos2017}).

Until now, the recommendations by the PDG for the physical properties of these states were followed. Given that the application of the methodology of Ref.~\cite{matsinos2020} in case of the scalar-isoscalar and vector-isovector 
mesons resulted in occasional departures from the recommended values by the PDG, it seemed prudent to pursue the extraction of estimates for the relevant physical properties of the HBRs.

This paper will be kept as short as possible. Given that the method was described in Section 2 of Ref.~\cite{matsinos2020}, the reader is referred to that paper for all details regarding the statistical analysis of the data; 
as Ref.~\cite{matsinos2020} is publicly available, there is no need to repeat in this paper any of the material contained therein. Consequently, three sections suffice for the purposes of this work. The current section serves 
as introduction to the subject; the subsequent section details the results obtained from the statistical analysis of the data; and the last section contains the conclusions of this paper.

It must be borne in mind that, as in Ref.~\cite{matsinos2020}, all masses will be expressed in energy units, namely in MeV. The abbreviation MF will stand for `minimisation function'.

\section{\label{sec:Results}Results}

Before entering the details, one comment on the two naming conventions of the HBRs is due. Justifying the shift from the old to the new notation in 2012, E.~Klempt and R.L.~Workman wrote in their note `$N$ and $\Delta$ 
resonances', preceding the relevant particle listings \cite{beringer2012}: ``In the past, when nearly all resonance information came from elastic $\pi N$ scattering, it was common to label resonances with the incoming partial 
wave $L_{2I, 2J}$, as in $\Delta (1232)$ $P_{33}$ and $N (1680)$ $F_{15}$. However, most recent information has come from $\gamma N$ experiments. Therefore, we have replaced $L_{2I, 2J}$ with the spin-parity $J^P$ of the state, 
as in $\Delta (1232)$ $3/2^+$ and $N (1680)$ $5/2^+$.''

In this work, both notations will be given in Tables \ref{tab:NBaryons} and \ref{tab:DeltaBaryons}. The `translation' from the new to the old notation is straightforward. For the sake of example, the $N (1520)$ resonance with 
$I (J^P) = 1/2 (3/2^-)$ carries the subscript `13' in the old notation and, as $J = L \pm S=3/2$ may be obtained from $L = 3/2 - 1/2 = 1$ or from $L = 3/2 + 1/2 = 2$, the state can be either $P$ or $D$. As the parity $P = (-1)^{L+1}$ 
of the state is negative, $L$ is even: therefore, the old notation for the $N (1520)$ resonance is $D_{13} (1520)$.

The large range of the uncertainties, accompanying the input datapoints in some of the datasets of this work, along with the general sparseness and the admittedly poor quality (discrepant input) of the information on the 
physical properties of the HBRs, necessitate changes (compared to Ref.~\cite{pdg2020}) to the treatment of the input data. Regarding the former issue (range of uncertainties), one example is revealing. Two of the available 
datapoints for the branching fraction of the $N(1535)$ to $\pi N$ final states - namely those corresponding to the smallest and largest uncertainties - are given in Ref.~\cite{pdg2020} as: $35.5(2)~\%$ (ARNDT06) and $50(10)~\%$ 
(CUTKOSKY80). If the two reported uncertainties are taken at face value, and these datapoints (as they appear in Ref.~\cite{pdg2020}) are submitted to the optimisation, then the weight of the first entry (in the calculation) 
will be $2\,500$ times that of the second. To start with, such a disparity does not tally with the requirement for robustness. As a matter of fact, when one performs that particular one-parameter fit to the eleven available 
datapoints (two of which are the aforementioned observations), the fit is dragged towards the ARNDT06 result (regardless of the choice of the MF) and the resulting uncertainty of the model parameter comes out of the fit 
unacceptably small, mildly exceeding the uncertainty of the datapoint carrying the largest weight (and conveying a spurious impression of a highly precise result). Not only is such a result hardly representative of the range 
of values in the input data, it also rests upon the correctness of just one input value. Without doubt, precision in measurements is widely appreciated, however only if complemented by reliability (which is hard to ascertain 
when only one precise measurement is available). If it so happens that future measurements demonstrate that a highly influential datapoint (used in a calculation) had been erroneous, then all results, obtained from databases 
containing that particular datapoint, are subject to sizeable adjustment.

A second problem with the data used herein is that the datasets frequently contain discrepant input. One glaring example is the datafile relating to the branching fraction of the $N(1900)$ to $\pi N$ final states, which 
includes (among other observations) the datapoints: $1.9(1)~\%$ (HUNT19) and $25(1)~\%$ (SHKLYAR13), i.e., two results which are separated by about $23 \sigma$. There can be no doubt that \emph{at least} one of these two 
observations is in serious error.

Like many, I detest the manipulation of experimental data. Nevertheless, I see no alternative in the problem treated herein, other than attempting to reduce the influence of the datapoints with too small (when compared to 
the remaining datapoints of a given dataset) uncertainties. To provide a meaningful solution, generally applicable to all input datasets, I will put forward the following scheme. Prior to the submission of the input datasets 
to the optimisation, the median value of the elements in each array of input uncertainties will be obtained, and will replace those of the uncertainties of the input datapoints which fall short of that median value. The 
advantage of this scheme (extraction of more robust results) outweighs what would appear to some as disadvantages, namely the loss in precision and the extraction of a larger fitted uncertainty of the constant achieving the 
optimal description of the input datapoints (i.e., of the model parameter $\avg{y}$ in the language of Ref.~\cite{matsinos2020}). I thus choose to err on the side of caution.

As in Ref.~\cite{matsinos2020}, the recommendations of this work will rest upon the entirety of the observations listed in Ref.~\cite{pdg2020}, as those results appear therein. No attempt will be made to clarify which of 
these measurements may be considered to be independent observations and which not (see also comments in the introduction of Section 3 in Ref.~\cite{matsinos2020}, just before Section 3.1 therein). In the subject treated in 
this work, all recommended values by the PDG are what they call `ESTIMATES': they have been ``based on the observed range of the data'' and have not been obtained ``from a formal statistical procedure'' \cite{pdg2020}.

\subsection{\label{sec:NBaryons}The $N$ HBRs}

A total of $229$ observations are available in Ref.~\cite{pdg2020} for the seven $N$ HBRs with masses below $2$ GeV. The recommended values by the PDG have been based on $115$ of these datapoints, twelve of which will be 
identified as outliers herein.

Table \ref{tab:NBaryons} and Figs.~\ref{fig:MassNBaryons}-\ref{fig:BFNBaryons} contain the results of this work for the $N$ HBRs: they correspond to the optimisation featuring a MF using Andrews weights \cite{matsinos2020}. 
The optimisation with Tukey weights was also performed in all cases, for verification of the results, yielding almost identical output (which is hardly a surprise). Given the sparseness of the available data, the processing 
of the `trimmed' datasets (i.e., of the datasets which are devoid of the outliers obtained from the aforementioned two robust-optimisation methods), in terms of the standard $\chi^2$ MF or of the method featuring the Cumulative 
Distribution Function \cite{matsinos2020}, was not performed.

\begin{table}
{\bf \caption{\label{tab:NBaryons}}}Summary of the recommended values for the physical properties of the $N$ ($S=0,I=1/2$) HBRs below $2$ GeV. The masses and the total decay widths are expressed in MeV, the branching fractions 
in percent.
\vspace{0.3cm}
\begin{center}
\begin{tabular}{|c|c|c|}
\hline
Physical quantity & PDG \cite{pdg2020} & This work\\
\hline
\hline
\multicolumn{3}{|c|}{$N (1440)$ $1/2^+$; $P_{11}$}\\
\hline
Breit-Wigner mass & $1410-1470$, $\approx 1440$ & $1422.2^{+9.5}_{-9.4}$\\
Breit-Wigner width & $250-450$, $\approx 350$ & $314^{+48}_{-46}$\\
Branching fraction to $\pi N$ & $55-75$, $\approx 65$ & $60.7^{+2.7}_{-2.6}$\\
\hline
\multicolumn{3}{|c|}{$N (1535)$ $1/2^-$; $S_{11}$}\\
\hline
Breit-Wigner mass & $1515-1545$, $\approx 1530$ & $1527.2^{+5.0}_{-4.2}$\\
Breit-Wigner width & $125-175$, $\approx 150$ & $137.0^{+7.4}_{-7.3}$\\
Branching fraction to $\pi N$ & $32-52$, $\approx 42$ & $39.0 \pm 2.4$\\
\hline
\multicolumn{3}{|c|}{$N (1650)$ $1/2^-$; $S_{11}$}\\
\hline
Breit-Wigner mass & $1635-1665$, $\approx 1650$ & $1661.1^{+3.8}_{-3.9}$\\
Breit-Wigner width & $100-150$, $\approx 125$ & $130.0^{+9.0}_{-8.8}$\\
Branching fraction to $\pi N$ & $50-70$, $\approx 60$ & $63.8^{+8.5}_{-7.8}$\\
\hline
\multicolumn{3}{|c|}{$N (1710)$ $1/2^+$; $P_{11}$}\\
\hline
Breit-Wigner mass & $1680-1740$, $\approx 1710$ & $1724 \pm 12$\\
Breit-Wigner width & $80-200$, $\approx 140$ & $155 \pm 22$\\
Branching fraction to $\pi N$ & $5-20$, $\approx 10$ & $9.2^{+2.7}_{-2.6}$\\
\hline
\multicolumn{3}{|c|}{$N (1720)$ $3/2^+$; $P_{13}$}\\
\hline
Breit-Wigner mass & $1680-1750$, $\approx 1720$ & $1714.9 \pm 8.0$\\
Breit-Wigner width & $150-400$, $\approx 250$ & $196 \pm 23$\\
Branching fraction to $\pi N$ & $8-14$, $\approx 11$ & $14.1 \pm 1.4$\\
\hline
\end{tabular}
\end{center}
\vspace{0.5cm}
\end{table}

\begin{table*}
{\bf Table \ref{tab:NBaryons} continued}
\vspace{0.3cm}
\begin{center}
\begin{tabular}{|c|c|c|}
\hline
Physical quantity & PDG \cite{pdg2020} & This work\\
\hline
\hline
\multicolumn{3}{|c|}{$N (1895)$ $1/2^-$; $S_{11}$}\\
\hline
Breit-Wigner mass & $1870-1920$, $\approx 1895$ & $1890^{+22}_{-23}$\\
Breit-Wigner width & $80-200$, $\approx 120$ & $440^{+89}_{-91}$\\
Branching fraction to $\pi N$ & $2-18$, $\approx 10$ & $10.5^{+7.3}_{-6.9}$\\
\hline
\multicolumn{3}{|c|}{$N (1900)$ $3/2^+$; $P_{13}$}\\
\hline
Breit-Wigner mass & $1890-1950$, $\approx 1920$ & $1917 \pm 15$\\
Breit-Wigner width & $100-320$, $\approx 200$ & $298^{+49}_{-50}$\\
Branching fraction to $\pi N$ & $2-20$, $\approx 10$ & $3.0 \pm 1.9$\\
\hline
\end{tabular}
\end{center}
\vspace{0.5cm}
\end{table*}

At this point, a few remarks are due.
\begin{enumerate}
\item Out of the $229$ observations, $20$ were identified as outliers by the two robust-optimisation methods of this work. As it might be useful in other works, the list of the outliers will be given. Regarding this list, 
the flags `used' and `not used' indicate whether or not the relevant observation(s) was/were used in Ref.~\cite{pdg2020}.
\begin{itemize}
\item SHKLYAR13 and ARNDT06 (used), PENNER02C (not used) from the dataset corresponding to the Breit-Wigner mass of the $N(1440)$;
\item HOEHLER79 (used), PENNER02C (not used) from the dataset corresponding to the Breit-Wigner width of the $N(1440)$;
\item ARNDT06 (used) from the dataset corresponding to the branching fraction of the $N(1440)$ to $\pi N$ final states;
\item ARNDT06 (used), ARNDT04 (not used) from the dataset corresponding to the Breit-Wigner mass of the $N(1535)$;
\item KASHEVAROV17 and ARNDT06 (used) from the dataset corresponding to the Breit-Wigner mass of the $N(1650)$;
\item BATINIC10 (not used) from the dataset corresponding to the Breit-Wigner width of the $N(1650)$;
\item HUNT19 (used) from the dataset corresponding to the Breit-Wigner mass of the $N(1710)$;
\item KASHEVAROV17, SOKHOYAN15A, and HOEHLER79 (used), ANISOVICH12A (not used) from the dataset corresponding to the Breit-Wigner width of the $N(1895)$;
\item SHRESTHA12A and PENNER02C (not used) from the dataset corresponding to the Breit-Wigner width of the $N(1900)$; and
\item SHKLYAR13 (used), PENNER02C (not used) from the dataset corresponding to the branching fraction of the $N(1900)$ to $\pi N$ final states.
\end{itemize}
\item One case stands out and is worth commenting on. Regarding the Breit-Wigner width of the $N(1895)$, the nine available observations cluster in two distinct (well-separated) sets: four observations suggest a low value 
of the width of that state (unweighted average: $\approx 109$ MeV), whereas five results point to a high value (unweighted average: $\approx 394$ MeV). The recommended value by the PDG is low (see Table \ref{tab:NBaryons} 
and Fig.~\ref{fig:WidthNBaryons}) because it has been based on three observations belonging to the former set and only one (the HUNT19 result in the notation of Ref.~\cite{pdg2020}) from the latter. (Incidentally, it is 
unclear how the HUNT19 result enters their recommendation at all!) On the other hand, both robust-optimisation methods of this work flag all four low observations as outliers, thus resulting in a high estimate for the 
Breit-Wigner width of the $N(1895)$.
\item This work suggests that the branching fraction of the $N(1990)$ to $\pi N$ final states is small, about $3~\%$. The two available high estimates, i.e., $25(1)~\%$ and $16(2)~\%$, are identified as outliers by both 
robust-optimisation methods.
\item Inspection of Table \ref{tab:NBaryons} and of Figs.~\ref{fig:MassNBaryons}-\ref{fig:BFNBaryons} demonstrates that - excepting the aforementioned problem regarding the Breit-Wigner width of the $N(1895)$ - the agreement 
between the two sets of recommendations, i.e., the ones by the PDG and those from this work, is satisfactory. The results of this work are more precise on almost all occasions. To resolve the obvious disagreement in case 
of the Breit-Wigner width of the $N(1895)$, additional data are required.
\end{enumerate}

\subsection{\label{sec:DeltaBaryons}The $\Delta$ HBRs}

A total of $80$ observations are available in Ref.~\cite{pdg2020} for the three $\Delta$ HBRs with masses below $2$ GeV. The recommended values by the PDG are based on $44$ observations, three of which will be identified as 
outliers herein.

Table \ref{tab:DeltaBaryons} and Figs.~\ref{fig:MassDeltaBaryons}-\ref{fig:BFDeltaBaryons} contain the results of this work for the $\Delta$ HBRs. The agreement between the two sets of recommendations is satisfactory in all cases. 
Only two comments are worth making. This work recommends a Breit-Wigner mass for the $\Delta (1600)$ which is in the vicinity of the high end of the range recommended by the PDG. It also suggests that the branching fraction 
to $\pi N$ final states of the $\Delta (1900)$ is close to the low end of the PDG range for that quantity.

\begin{table}
{\bf \caption{\label{tab:DeltaBaryons}}}The equivalent of Table \ref{tab:NBaryons} for the physical properties of the $\Delta$ ($S=0,I=3/2$) HBRs below $2$ GeV. In the context of the ETH model \cite{goudsmit1994,matsinos2017}, 
the contributions from the $s$- and $u$-channel graphs with a $\Delta (1232)$ intermediate state are included in the main contributions shown in Fig.~\ref{fig:FeynmanGraphsETHZ}. As a result, the $\Delta (1232)$ does not 
belong to the set of $\Delta$ HBRs; the results of the fits to the available data in case of the $\Delta (1232)$ are given separately in Appendix \ref{App:AppA}.
\vspace{0.3cm}
\begin{center}
\begin{tabular}{|c|c|c|}
\hline
Physical quantity & PDG \cite{pdg2020} & This work\\
\hline
\hline
\multicolumn{3}{|c|}{$\Delta (1600)$ $3/2^+$; $P_{33}$}\\
\hline
Breit-Wigner mass & $1500-1640$, $\approx 1570$ & $1654 \pm 29$\\
Breit-Wigner width & $200-300$, $\approx 250$ & $255^{+42}_{-41}$\\
Branching fraction to $\pi N$ & $8-24$, $\approx 16$ & $14.0 \pm 2.4$\\
\hline
\multicolumn{3}{|c|}{$\Delta (1620)$ $1/2^-$; $S_{31}$}\\
\hline
Breit-Wigner mass & $1590-1630$, $\approx 1610$ & $1605.3^{+5.7}_{-5.6}$\\
Breit-Wigner width & $110-150$, $\approx 130$ & $130.8^{+9.1}_{-8.9}$\\
Branching fraction to $\pi N$ & $25-35$, $\approx 30$ & $29.8^{+2.2}_{-2.1}$\\
\hline
\multicolumn{3}{|c|}{$\Delta (1910)$ $1/2^+$; $P_{31}$}\\
\hline
Breit-Wigner mass & $1850-1950$, $\approx 1900$ & $1875^{+32}_{-33}$\\
Breit-Wigner width & $200-400$, $\approx 300$ & $291^{+38}_{-37}$\\
Branching fraction to $\pi N$ & $15-30$, $\approx 20$ & $15.1^{+2.1}_{-2.0}$\\
\hline
\end{tabular}
\end{center}
\vspace{0.5cm}
\end{table}

It might be useful in other works to list the six identified outliers in the case of the $\Delta$ HBRs; they are: SOKHOYAN15A and HOEHLER79 (used), ANISOVICH12A (not used) from the dataset corresponding to the Breit-Wigner 
mass of the $\Delta (1600)$; PENNER02C (not used) from the dataset corresponding to the Breit-Wigner width of the $\Delta (1620)$; and ARNDT06 (used), VRANA00 (not used) from the dataset corresponding to the Breit-Wigner 
mass of the $\Delta (1910)$. Again, the flags `used' and `not used' indicate whether or not the observations were used in obtaining the recommended values in Ref.~\cite{pdg2020}.

\section{\label{sec:Conclusions}Conclusions}

Estimates for the Breit-Wigner masses and for the partial decays widths to pion-nucleon ($\pi N$) final states of the well-established (four-star) $s$ and $p$ baryon resonances below $2$ GeV were obtained in this work on the 
basis of the entirety of the available data, as they appear in the recent compilation by the Particle Data Group (PDG) \cite{pdg2020}. These physical properties are interesting in the context of a $\pi N$ interaction model 
\cite{matsinos2017}, based on the contributions of $s$-, $u$-, and $t$-channel Feynman graphs to the hadronic part of the $s$- and $p$-wave scattering amplitudes \cite{goudsmit1994}.

On almost all occasions, the results of this work are more precise than the ranges of probable values, recommended by the PDG. However, this might not be surprising, given that the relevant PDG recommendations have not been 
obtained from an appropriate statistical analysis of the available data.

Regarding the $N$ ($S=0,I=1/2$) baryons, the agreement between the recommended values by the PDG and the results of this work is satisfactory, save for the Breit-Wigner width of the $N(1895)$ state. This disagreement is due 
to the selection of the input: the available datapoints may be categorised in terms of two distinct sets of values, one centred around $110$ MeV (low estimates), the other around $390$ MeV (high estimates). Evidently, 
additional data is required in order to resolve this discrepancy.

Finally, regarding the $\Delta$ ($S=0,I=3/2$) baryons, the agreement between the recommended values by the PDG and the results of this work is satisfactory in all cases.

\begin{ack}
The Feynman graphs of this paper have been drawn with the software package JaxoDraw \cite{binosi2004,binosi2009}, available from jaxodraw.sourceforge.net. The remaining figures have been created with MATLAB$^{\textregistered}$~(The 
MathWorks, Inc., Natick, Massachusetts, United States).
\end{ack}

\clearpage
\begin{figure}
\begin{center}
\includegraphics [width=15.5cm] {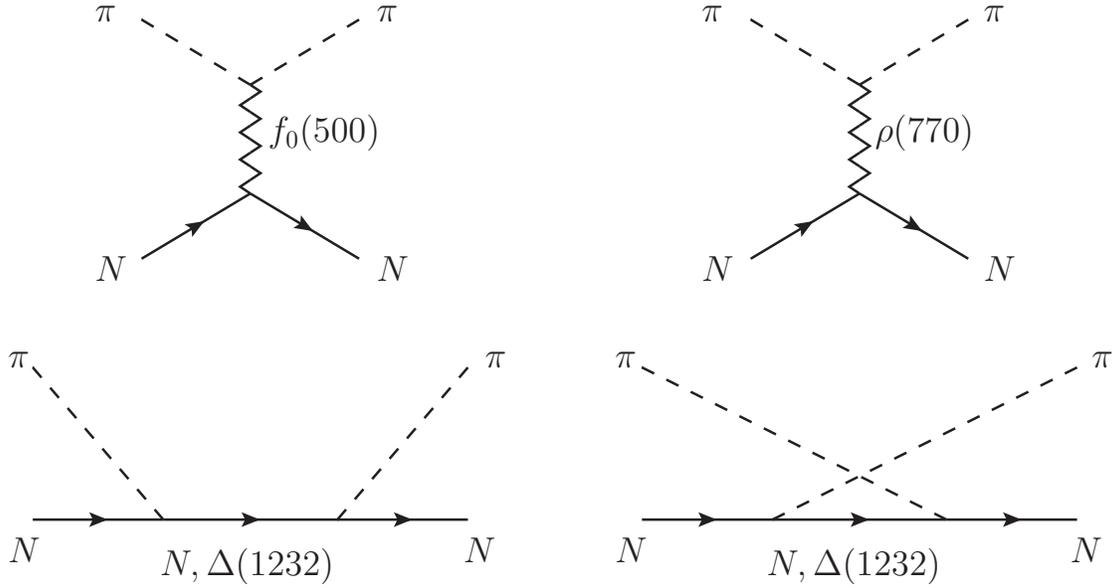}
\caption{\label{fig:FeynmanGraphsETHZ}The main Feynman graphs of the ETH model: scalar-isoscalar $I^G\,(J^{PC}) = 0^+\,(0^{++})$ and vector-isovector $I^G\,(J^{PC}) = 1^+\,(1^{--})$ $t$-channel graphs (upper part), and $N$ 
and $\Delta (1232)$ $s$- and $u$-channel graphs (lower part). Not shown in this figure, but also analytically included in the model, are the small contributions from all other scalar-isoscalar and vector-isovector mesons 
with rest masses below $2$ GeV, as well as from all well-established (four-star) $s$ and $p$ higher baryon resonances (HBRs) in the same mass range.}
\end{center}
\end{figure}

\begin{figure}
\begin{center}
\includegraphics [width=15.5cm] {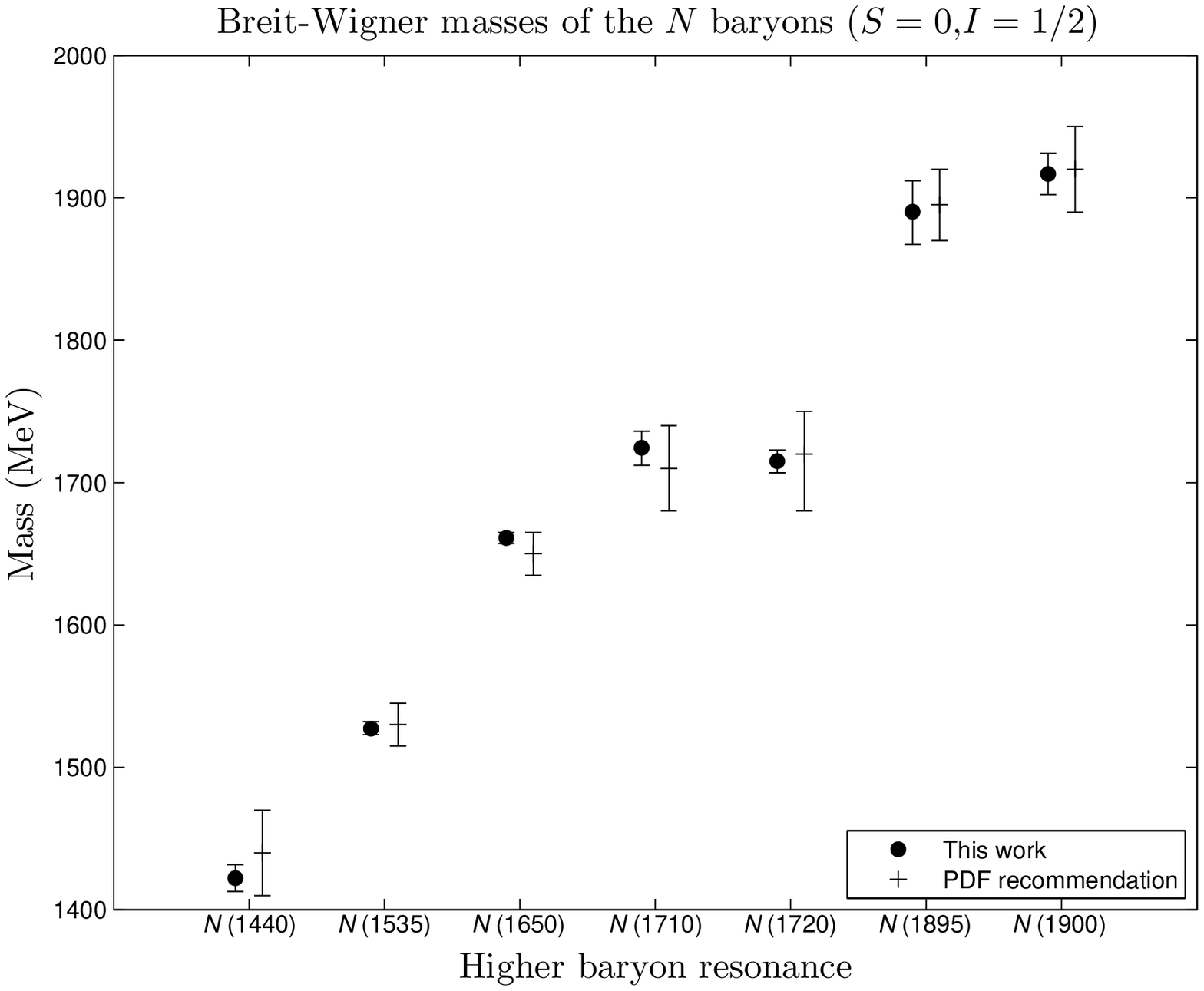}
\caption{\label{fig:MassNBaryons}The results of this work for the Breit-Wigner masses of the seven well-established (four-star) $N$ HBRs below $2$ GeV (see caption of Fig.~\ref{fig:FeynmanGraphsETHZ}), along with the 
recommended values by the PDG. The uncertainties of the results of this work correspond to $1 \sigma$ in the normal distribution. Each PDG recommendation is marked by a `plus' sign, presumably indicating the most probable 
value, whereas their uncertainties indicate an expected range of variation; it is unclear from Ref.~\cite{pdg2020} what this range represents in terms of quartiles in each distribution.}
\end{center}
\end{figure}

\begin{figure}
\begin{center}
\includegraphics [width=15.5cm] {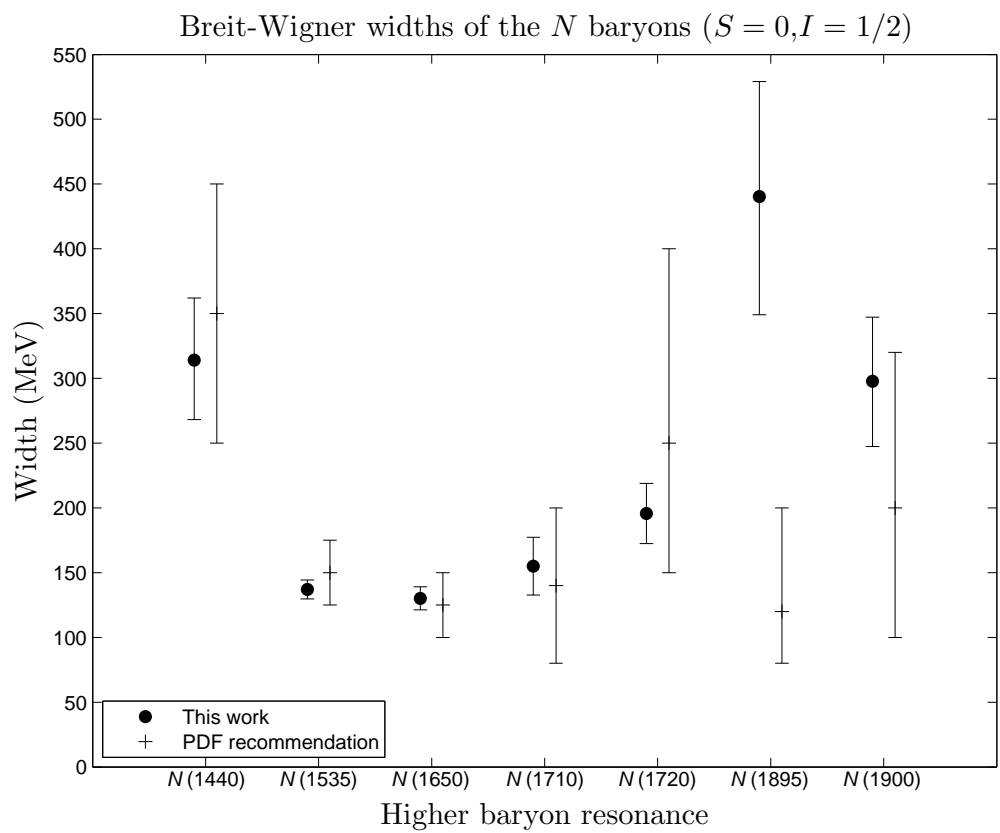}
\caption{\label{fig:WidthNBaryons}The equivalent of Fig.~\ref{fig:MassNBaryons} for the Breit-Wigner widths of the seven $N$ HBRs.}
\end{center}
\end{figure}

\begin{figure}
\begin{center}
\includegraphics [width=15.5cm] {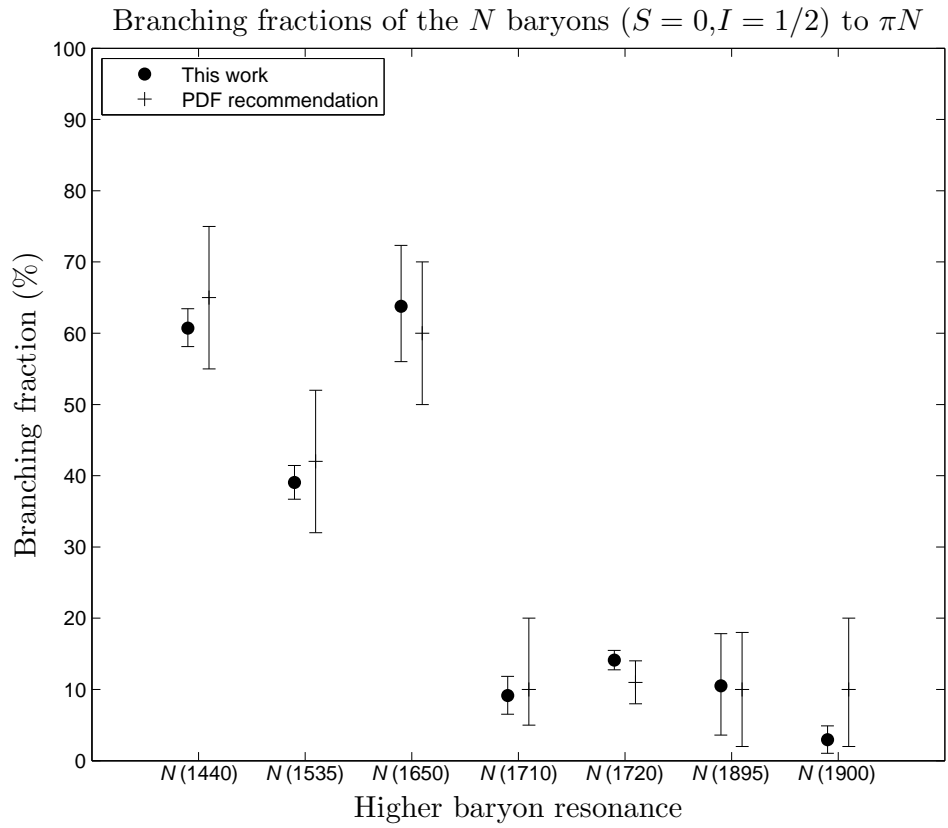}
\caption{\label{fig:BFNBaryons}The equivalent of Fig.~\ref{fig:MassNBaryons} for the branching fractions of the seven $N$ HBRs to $\pi N$ final states.}
\end{center}
\end{figure}

\begin{figure}
\begin{center}
\includegraphics [width=15.5cm] {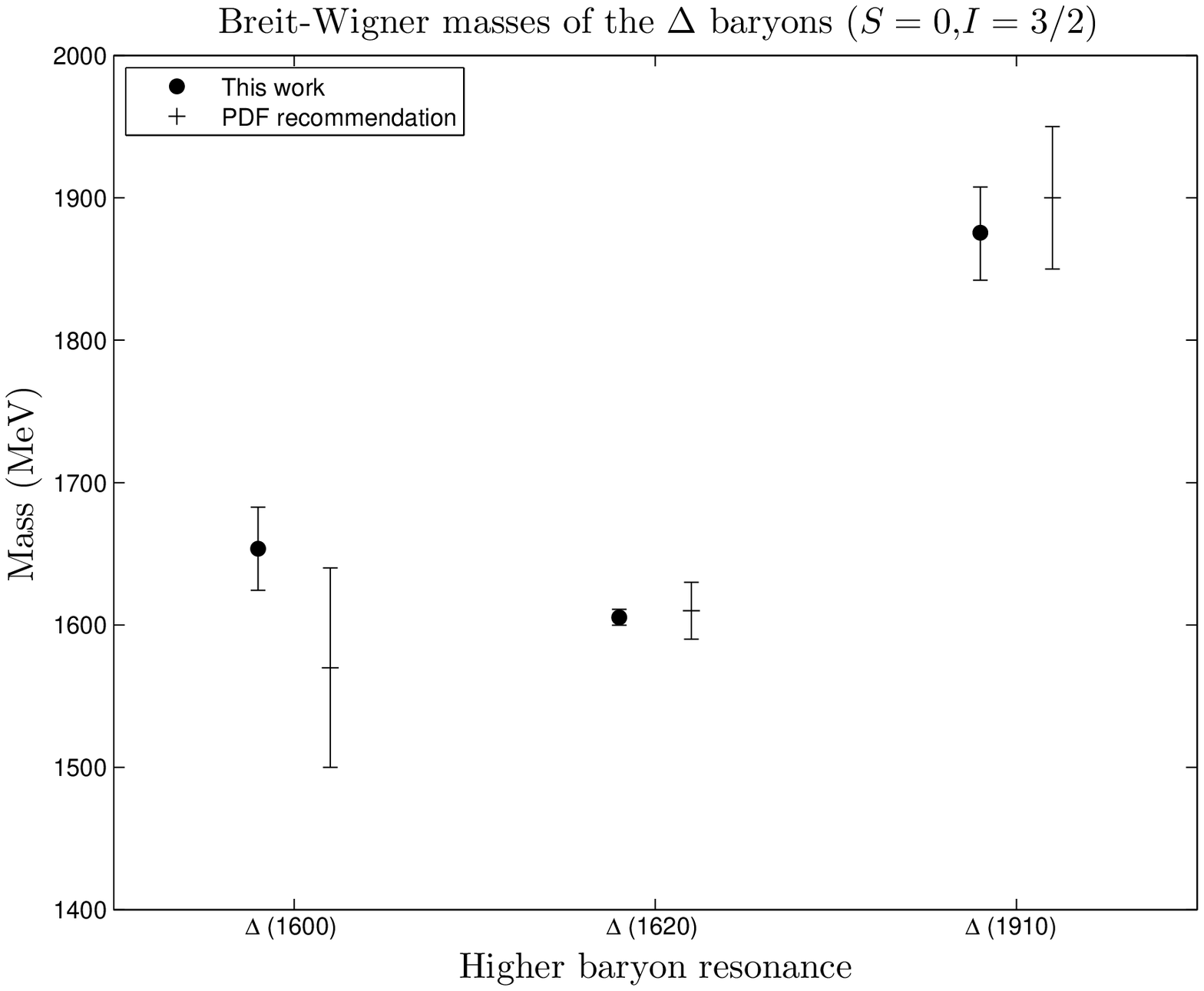}
\caption{\label{fig:MassDeltaBaryons}The results of this work for the Breit-Wigner masses of the three well-established (four-star) $\Delta$ HBRs below $2$ GeV (see caption of Fig.~\ref{fig:FeynmanGraphsETHZ}), along with the 
recommended values by the PDG. The uncertainties of the results of this work correspond to $1 \sigma$ in the normal distribution. Each PDG recommendation is marked by a `plus' sign, presumably indicating the most probable 
value, whereas their uncertainties indicate an expected range of variation; it is unclear from Ref.~\cite{pdg2020} what this range represents in terms of quartiles in each distribution.}
\end{center}
\end{figure}

\begin{figure}
\begin{center}
\includegraphics [width=15.5cm] {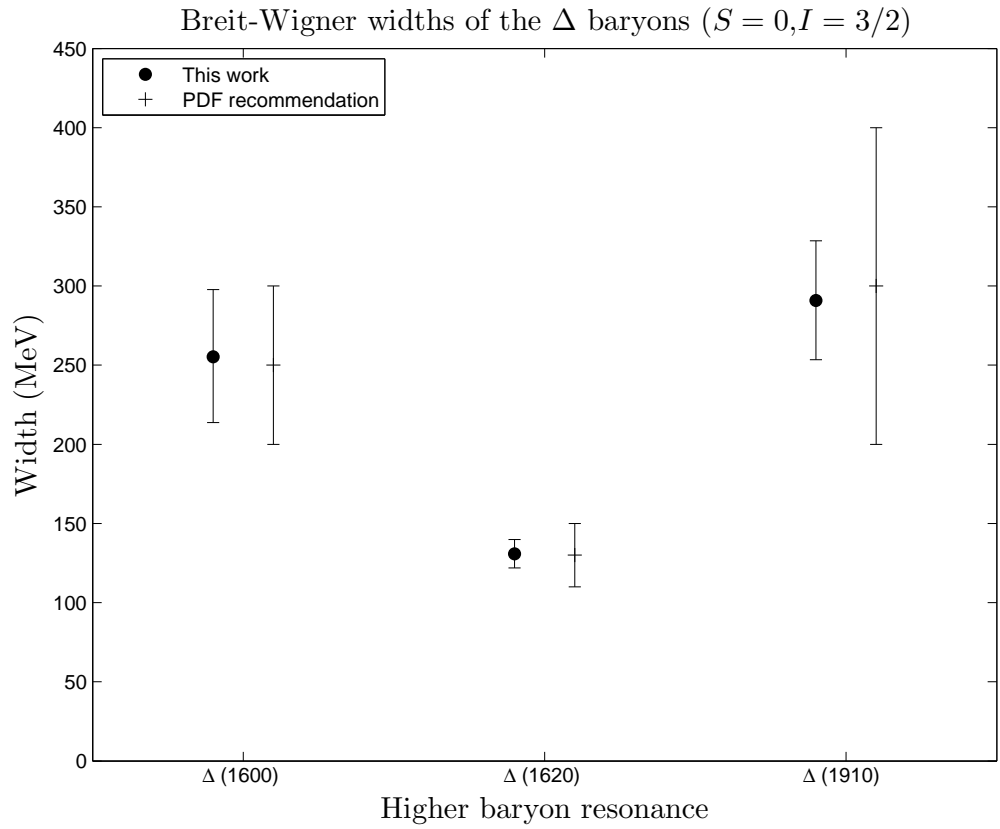}
\caption{\label{fig:WidthDeltaBaryons}The equivalent of Fig.~\ref{fig:MassDeltaBaryons} for the Breit-Wigner widths of the three $\Delta$ HBRs.}
\end{center}
\end{figure}

\begin{figure}
\begin{center}
\includegraphics [width=15.5cm] {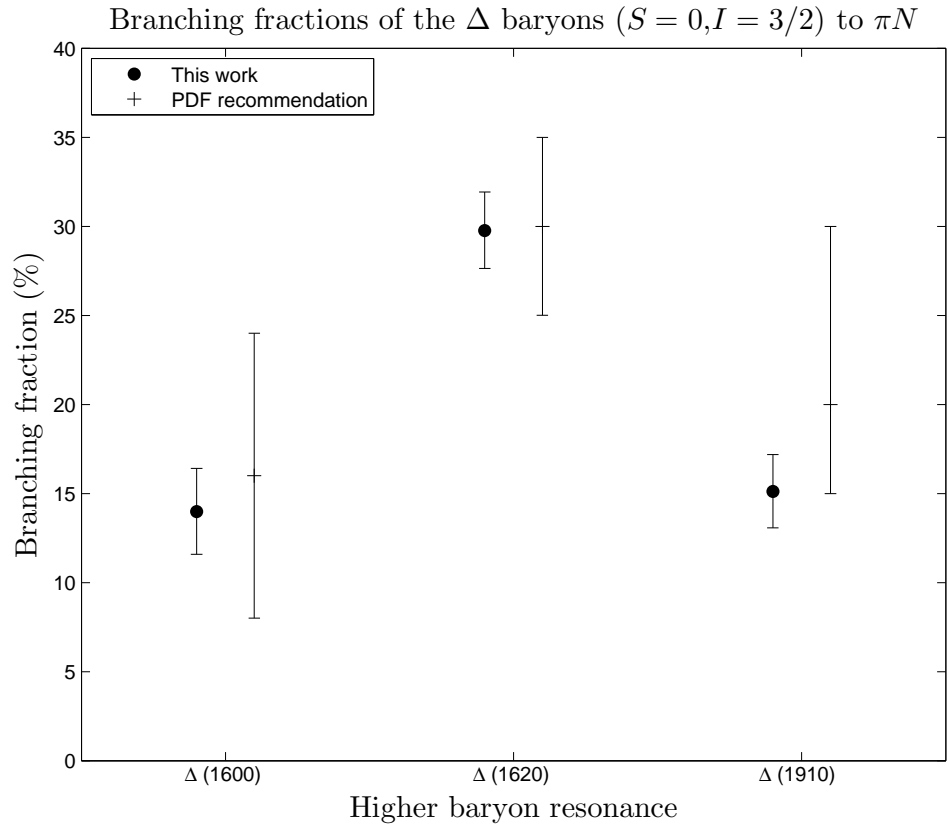}
\caption{\label{fig:BFDeltaBaryons}The equivalent of Fig.~\ref{fig:MassDeltaBaryons} for the branching fractions of the three $\Delta$ HBRs to $\pi N$ final states.}
\end{center}
\end{figure}

\clearpage
\newpage
\appendix
\section{\label{App:AppA}The $\Delta (1232)$ resonance}

In the context of the ETH model, this resonance does not belong to the HBRs: its $s$- and $u$-channel contributions to the hadronic part of the $s$- and $p$-wave scattering amplitudes are contained in the main input to the 
ETH model, see Fig.~\ref{fig:FeynmanGraphsETHZ}, lower part. For the sake of completeness, the methodology of this work was also applied to the data for this resonance, as listed in Ref.~\cite{pdg2020}, and yielded the results 
of Table \ref{tab:Delta1232}. Regarding the branching fraction of the $\Delta (1232)$ resonance to $\pi N$ final states, only one proper result (central value, accompanied by a meaningful uncertainty) is available, namely 
$99.39(1)~\%$, labelled in Ref.~\cite{pdg2020} as HUNT19. The lack of a result from this work implies that a statistical analysis in that case is impossible, and that the PDG recommendation is accepted. Finally, it is worth 
mentioning that the two robust-optimisation methods of this work did not identify any outliers among the input datapoints relevant to the $\Delta (1232)$ resonance.

\begin{table}
{\bf \caption{\label{tab:Delta1232}}}The equivalent of Table \ref{tab:DeltaBaryons} for the physical properties of the $\Delta (1232)$. In the context of the ETH model, this resonance does not belong to the set of the HBRs; 
its $s$- and $u$-channel contributions to the hadronic part of the $s$- and $p$-wave scattering amplitudes are taken into account in the main contributions shown in Fig.~\ref{fig:FeynmanGraphsETHZ}.
\vspace{0.3cm}
\begin{center}
\begin{tabular}{|c|c|c|}
\hline
Physical quantity & PDG \cite{pdg2020} & This work\\
\hline
\hline
\multicolumn{3}{|c|}{$\Delta (1232)$ $3/2^+$; $P_{33}$}\\
\hline
Breit-Wigner mass & $1230-1234$, $\approx 1232$ & $1231.33^{+0.91}_{-0.92}$\\
Breit-Wigner width & $114-120$, $\approx 117$ & $114.0^{+2.8}_{-2.9}$\\
Branching fraction to $\pi N$ & $99.4$ & $-$\\
\hline
\end{tabular}
\end{center}
\vspace{0.5cm}
\end{table}

\end{document}